\documentclass[12pt]{article}
\usepackage{graphics,graphicx}
\usepackage{amssymb,epsfig,amsmath,euscript,array}
\usepackage{cite}
\usepackage{pstricks}
\usepackage{color}

\makeatletter
\@addtoreset{equation}{section}
\makeatother

\newcounter{multieqs}

\newcommand{\be}{\begin{equation}}
\newcommand{\ee}{\end{equation}}

\newcommand{\bm}[1]{\mbox{\boldmath $#1$}}

\newcommand{\kslash}{k \!\!\! / }

\newcommand{\lslash}{l \!\! / }
\newcommand{\Pslash}{P \!\!\!\! / }

\newcommand{\islash}{i \!\!\! / }
\newcommand{\jslash}{j \!\!\! / }
\newcommand{\aslash}{a \!\!\! / }
\newcommand{\bslash}{{b \hspace{-6pt} \slash} }

\newcommand{\onslash}{1 \!\!\! / }
\newcommand{\twslash}{2 \!\!\!/ }
\newcommand{\thslash}{3 \!\!\!/ }
\newcommand{\foslash}{4 \!\!\! / }
\newcommand{\fislash}{5 \!\!\! / }

\newcommand{\mslash}{m \!\!\! / }

\def\bd{\begin{document}}
\def\ed{\end{document}}
\def\nn{\nonumber}
\def\bea{\begin{eqnarray}}
\def\eea{\end{eqnarray}}

\def\ab{(ijab)}
\def\ba{(ijba)}
\def\ijab{{\tr}_{+}(\islash\, \jslash\, \aslash \, \bslash)}
\def\ijba{{\tr}_{+}(\islash\, \jslash\, \bslash \, \aslash)}
\def\ijaP{{\tr}_{+}(\islash\, \jslash\, \aslash \, \Pslash)}
\def\ijPLa{{\tr}_{+}(\islash\, \jslash\, \Pslash_L \, \aslash)}
\def\ijaPL{{\tr}_{+}(\islash\, \jslash\, \aslash \, \Pslash_L)}
\def\ijPLza{{\tr}_{+}(\islash\, \jslash\, \Pslash_{L;z} \, \aslash)}
\def\ijaPLz{{\tr}_{+}(\islash\, \jslash\, \aslash \, \Pslash_{L;z})}
\def\ijPa{{\tr}_{+}(\islash\, \jslash\, \Pslash \, \aslash)}
\def\iaPb{{\tr}_{+}(\islash\, \aslash\, \Pslash \, \bslash)}
\def\ibPa{{\tr}_{+}(\islash\, \bslash\, \Pslash \, \aslash)}
\def\ijPmu{{\tr}_{+}(\islash\, \jslash\, \Pslash \, \mu)}
\def\ibmuP{{\tr}_{+}(\islash\, \bslash\, \mu \, \Pslash)}
\def\ibmua{{\tr}_{+}(\islash\, \bslash\, \mu \, \aslash)}
\def\iamub{{\tr}_{+}(\islash\, \aslash\, \mu \, \bslash)}
\def\jaPb{{\tr}_{+}(\jslash\, \aslash\, \Pslash \, \bslash)}
\def\ijmuP{{\tr}_{+}(\islash\, \jslash\, \mu \, \Pslash)}
\def\ijmum{{\tr}_{+}(\islash\, \jslash\, \mu \, \mslash)}
\def\ijmmu{{\tr}_{+}(\islash\, \jslash\, \mslash \, \mu)}
\def\ijmP{{\tr}_{+}(\islash\, \jslash\, \mslash \, \Pslash)}
\def\iabP{{\tr}_{+}(\islash\, \aslash\, \bslash \, \Pslash)}
\def\ijbP{{\tr}_{+}(\islash\, \jslash\, \bslash \, \Pslash)}
\def\jbPa{{\tr}_{+}(\jslash\, \bslash\, \Pslash \, \aslash)}
\def\ijPb{{\tr}_{+}(\islash\, \jslash\, \Pslash \, \bslash)}
\def\jbmua{{\tr}_{+}(\jslash\, \bslash\, \mu \, \aslash)}

\def\loablt{ {\tr}_{+}(\lslash_1\, \aslash \, \bslash\, \lslash_2)}

\def\ijlolt{{\tr}_{+}(\islash\, \jslash\, \lslash_1 \, \lslash_2)}
\def\ijltlo{{\tr}_{+}(\islash\, \jslash\, \lslash_2 \, \lslash_1)}
\def\ibloa{{\tr}_{+}(\islash\, \bslash\, \lslash_1 \, \aslash)}
\def\jaltb{{\tr}_{+}(\jslash\, \aslash\, \lslash_2 \, \bslash)}
\def\ialtb{{\tr}_{+}(\islash\, \aslash\, \lslash_2 \, \bslash)}
\def\bltloa{{\tr}_{+}(\bslash\, \lslash_2\, \lslash_1 \, \aslash)}
\def\jbloa{{\tr}_{+}(\jslash\, \bslash\, \lslash_1 \, \aslash)}
\def\ibPb{{\tr}_{+}(\islash\, \bslash\, \Pslash \, \bslash)}
\def\ijltb{{\tr}_{+}(\islash\, \jslash\, \lslash_2 \, \bslash)}

\def\ijloa{{\tr}_{+}(\islash\, \jslash\,  \lslash_1 \, \aslash)}
\def\ijblt{{\tr}_{+}(\islash\, \jslash\,  \bslash \, \lslash_2)}

\def\jakb{{\tr}_{+}(\jslash\, \aslash\, \kslash \, \bslash)}
\def\iakb{{\tr}_{+}(\islash\, \aslash\, \kslash \, \bslash)}

\def\tofo{{\tr}_{+}(\onslash\, \thslash\, \twslash \, \foslash)}
\def\foto{{\tr}_{+}(\onslash\, \thslash\, \foslash \, \twslash)}
\def\tofi{{\tr}_{+}(\onslash\, \thslash\, \twslash \, \fislash)}
\def\fito{{\tr}_{+}(\onslash\, \thslash\, \fislash \, \twslash)}

\def\lrangle#1#2{\langle #1\,#2\rangle}

\def\Li{{$\rm Li}_2$}
\def\eps{\epsilon}
\def\epsuv{{\epsilon_{\rm \mbox{\tiny UV}}}}
\let\bm=\bibitem
\let\la=\label

\def\npb#1#2#3{Nucl. Phys. {\bf{B#1}} #3 (#2)}
\def\plb#1#2#3{Phys. Lett. {\bf{#1B}} #3 (#2)}
\def\prl#1#2#3{Phys. Rev. Lett. {\bf{#1}} #3 (#2)}
\def\prd#1#2#3{Phys. Rev. {D \bf{#1}} #3 (#2)}
\def\cmp#1#2#3{Comm. Math. Phys. {\bf{#1}} #3 (#2)}
\def\cqg#1#2#3{Class. Quantum Grav. {\bf{#1}} #3 (#2)}
\def\nppsa#1#2#3{Nucl. Phys. B (Proc. Suppl.) {\bf{#1A}}#3 (#2)}
\def\ap#1#2#3{Ann. of Phys. {\bf{#1}} #3 (#2)}
\def\ijmp#1#2#3{Int. J. Mod. Phys. {\bf{A#1}} #3 (#2)}
\def\rmp#1#2#3{Rev. Mod. Phys. {\bf{#1}} #3 (#2)}
\def\mpla#1#2#3{Mod. Phys. Lett. {\bf A#1} #3 (#2)}
\def\jhep#1#2#3{J. High Energy Phys. {\bf #1} #3 (#2)}
\def\atmp#1#2#3{Adv. Theor. Math. Phys. {\bf #1} #3 (#2)}
\newcommand{\EQ}[1]{\begin{equation} #1 \end{equation}}
\newcommand{\AL}[1]{\begin{subequations}\begin{align} #1 \end{align}\end{subequations}}
\newcommand{\SP}[1]{\begin{equation}\begin{split} #1 \end{split}\end{equation}}
\newcommand{\ALAT}[2]{\begin{subequations}\begin{alignat}{#1} #2 \end{alignat}
                        \end{subequations}}
\def\beqa{\begin{eqnarray}}
\def\eeqa{\end{eqnarray}}
\def\beq{\begin{equation}}
\def\eeq{\end{equation}}
\def\sst{\scriptscriptstyle}
\def\thetabar{\bar\theta}
\def\Tr{{\rm Tr}}
\def\one{\mbox{1 \kern-.59em {\rm l}}}
 \def\Nh{\hat{N}}

\newcommand{\half}{{\textstyle {1 \over 2}}}

\def\a{\alpha}      \def\da{{\dot\alpha}}
\def\b{\beta}       \def\db{{\dot\beta}}
\def\c{\gamma}  \def\G{\Gamma}  \def\cdt{\dot\gamma}
\def\d{\delta}  \def\D{\Delta}  \def\ddt{\dot\delta}
\def\e{\epsilon}        \def\vare{\varepsilon}
\def\f{\phi}    \def\F{\Phi}    \def\vvf{\f}
\def\h{\eta}
\def\k{\kappa}
\def\l{\lambda} \def\L{\Lambda}
\def\m{\mu} \def\n{\nu}
\def\o{\omega}
\def\p{\pi} \def\P{\Pi}
\def\r{\rho}
\def\s{\sigma}  \def\S{\Sigma}
\def\t{\tau}
\def\th{\theta} \def\Th{\Theta} \def\vth{\vartheta}
\def\X{\Xeta}
\def\z{\zeta}
\def\de{\partial}

\def\cA{{\cal A}} \def\cB{{\cal B}} \def\cC{{\cal C}}
\def\cD{{\cal D}} \def\cE{{\cal E}} \def\cF{{\cal F}}
\def\cG{{\cal G}} \def\cH{{\cal H}} \def\cI{{\cal I}}
\def\cJ{{\cal J}} \def\cK{{\cal K}} \def\cL{{\cal L}}
\def\cM{{\cal M}} \def\cN{{\cal N}} \def\cO{{\cal O}}
\def\cP{{\cal P}} \def\cQ{{\cal Q}} \def\cR{{\cal R}}
\def\cS{{\cal S}} \def\cT{{\cal T}} \def\cU{{\cal U}}
\def\cV{{\cal V}} \def\cW{{\cal W}} \def\cX{{\cal X}}
\def\cY{{\cal Y}} \def\cZ{{\cal Z}}

\def\ua{\underline{\alpha}}
\def\ub{\underline{\phantom{\alpha}}\!\!\!\beta}
\def\uc{\underline{\phantom{\alpha}}\!\!\!\gamma}
\def\um{\underline{\mu}}
\def\ud{\underline\delta}
\def\ue{\underline\epsilon}
\def\una{\underline a}\def\unA{\underline A}
\def\unb{\underline b}\def\unB{\underline B}
\def\unc{\underline c}\def\unC{\underline C}
\def\und{\underline d}\def\unD{\underline D}
\def\une{\underline e}\def\unE{\underline E}
\def\unf{\underline{\phantom{e}}\!\!\!\! f}\def\unF{\underline F}
\def\unm{\underline m}\def\unM{\underline M}
\def\unn{\underline n}\def\unN{\underline N}
\def\unp{\underline{\phantom{a}}\!\!\! p}\def\unP{\underline P}
\def\unq{\underline{\phantom{a}}\!\!\! q}
\def\unQ{\underline{\phantom{A}}\!\!\!\! Q}
\def\unH{\underline{H}}

\def\As {{A \hspace{-6.4pt} \slash}\;}
\def\bs {{b \hspace{-6.4pt} \slash}\;}
\def\Ds {{D \hspace{-6.4pt} \slash}\;}
\def\ds {{\del \hspace{-6.4pt} \slash}\;}
\def\ss {{\s \hspace{-6.4pt} \slash}\;}
\def\ks {{ k \hspace{-6.4pt} \slash}\;}
\def\ps {{p \hspace{-6.4pt} \slash}\;}
\def\pas {{{p_1} \hspace{-6.4pt} \slash}\;}
\def\pbs {{{p_2} \hspace{-6.4pt} \slash}\;}
\def\Ps {{P \hspace{-6.4pt} \slash}\;}
\def\Qs {{Q \hspace{-6.4pt} \slash}\;}

\def\Fh{\hat{F}}
\def\Vh{\hat{V}}
\def\Xh{\hat{X}}
\def\ah{\hat{a}}
\def\xh{\hat{x}}
\def\yh{\hat{y}}
\def\ph{\hat{p}}
\def\xih{\hat{\xi}}
\def\psit{\tilde{\psi}}
\def\Psit{\tilde{\Psi}}
\def\tht{\tilde{\th}}
\def\lt{\tilde{\lambda}}
\def\hl{\hat{\lambda}}
\def\hlt{\hat{\tilde{\lambda}}}
\def\llt{\tilde{l}}
\def\At{\tilde{A}}
\def\Qt{\tilde{Q}}
\def\Rt{\tilde{R}}
\def\Nt{\tilde{N}}

\def\at{\tilde{a}}
\def\st{\tilde{s}}
\def\ft{\tilde{f}}
\def\pt{\tilde{p}}
\def\qt{\tilde{q}}
\def\vt{\tilde{v}}
\def\nt{\tilde{n}}

\def\delb{\bar{\partial}}
\def\bz{\bar{z}}
\def\bD{\bar{D}}
\def\bB{\bar{B}}

\def\bk{{\bf k}}
\def\bl{{\bf l}}
\def\bp{{\bf p}}
\def\bq{{\bf q}}
\def\br{{\bf r}}
\def\bx{{\bf x}}
\def\by{{\bf y}}
\def\bR{{\bf R}}
\def\bV{{\bf V}}

\def\d{\delta}\def\D{\Delta}\def\ddt{\dot\delta}
\def\pa{\partial} \def\del{\partial}
\def\xx{\times}
\def\uno{\mbox{1 \kern-.59em {\rm l}}}
\def\trp{^{\top}}
\def\inv{^{-1}}
\def\dag{{^{\dagger}}}
\def\pr{^{\prime}}
\def\lan{\langle}
\def\ran{\rangle}
\def\rar{\rightarrow}
\def\lar{\leftarrow}
\def\lrar{\leftrightarrow}
\newcommand{\0}{\,\!}      
\def\one{1\!\!1\,\,}
\def\im{\imath}
\def\jm{\jmath}
\newcommand{\tr}{\mbox{tr}}
\newcommand{\slsh}[1]{/ \!\!\!\! #1}
\def\vac{|0\rangle}
\def\lvac{\langle 0|}
\def\hlf{\frac{1}{2}}
\def\ove#1{\frac{1}{#1}}
\def\Box{\square}
\def\ZZ{\mathbb{Z}}
\def\CC#1{({\bf #1})}
\def\bcomment#1{}
\def\bfhat#1{{\bf \hat{#1}}}
\def\VEV#1{\left\langle #1\right\rangle}
\newcommand{\ex}[1]{{\rm e}^{#1}} \def\ii{{\rm i}}
\def\rr{{\rm r}} \def\rs{{\rm s}}\def\rv{{\rm v}}
\def\ri{{\rm i}}\def\rj{{\rm j}}
\newcommand{\lrbrk}[1]{\left(#1\right)}
\newcommand{\sfrac}[2]{{\textstyle\frac{#1}{#2}}}

\def\Li{{\rm Li}_2}

\font\mybb=msbm10 at 12pt
\def\bb#1{\hbox{\mybb#1}}

\font\myBB=msbm10 at 18pt
\def\BB#1{\hbox{\myBB#1}}

\begin{document}

\begin{flushright}
QMUL-PH-10-18
\end{flushright}

\vspace{20pt}

\begin{center}

{\Large \bf Form Factors in $\cN=4$ Super Yang-Mills       and  }
\\
\vspace{0.3cm}
{\Large \bf  Periodic Wilson Loops   }
\vspace{11pt}
\vspace{32pt}

{\mbox {\bf Andreas Brandhuber, Bill Spence,  Gabriele Travaglini and Gang Yang}}%
\footnote{
{\sffamily \{\tt a.brandhuber, w.j.spence, g.travaglini, g.yang\}@qmul.ac.uk }}

{\em Centre for Research in String Theory\\
School of Physics\\
Queen Mary University of London\\
Mile End Road, London, E1 4NS\\
United Kingdom
 }

\vspace{30pt} {\bf Abstract}

\end{center}

\noindent
We calculate form factors  of half-BPS operators in $\cN=4$ super Yang-Mills theory at tree level and one loop using   novel applications of recursion relations and unitarity.  In particular, we determine the expression of the one-loop form factors with two scalars and an arbitrary number of positive-helicity gluons. These quantities resemble closely  the MHV scattering amplitudes, including holomorphicity of the tree-level form factor, and the expansion in terms of two-mass easy
box functions of the one-loop result.
Next, we compare our result  for these form factors  to the calculation of a
particular periodic Wilson loop at one loop, finding agreement.
This suggests a novel duality  relating  form factors to periodic Wilson loops.

\setcounter{page}{0}
\thispagestyle{empty}
\newpage


\section{Introduction   }
\setcounter{footnote}{0}

One of the important challenges ahead is that of relating two apparently disconnected realms, that of scattering amplitudes, and that of correlation functions. The first candidate theory to study is naturally maximally supersymmetric Yang-Mills theory (SYM). 
A step in this direction was taken in \cite{Alday:2010zy, Eden:2010zz, Eden:2010ce}, where correlation functions of composite operators in $\cN=4$ SYM theory were  considered in certain special lightlike limits where the distances between adjacent insertion points become null.
Surprisingly, it was found that in the lightlike limit considered there  is a simple relation between
$(n+l)$-point correlators with  $l$ insertions of the Lagrangian, and the integrand of the $n$-point MHV amplitude evaluated
at $l$ loops.

In this paper we will focus on an interesting class of physical observables  which sit between the two worlds of amplitudes and correlation functions. These are  the form factors, i.e.~matrix elements of  gauge-invariant, composite operators $\cO$ between the vacuum and some external scattering state,%
\footnote{We note that the dimension of the form factor in \eqref{f1}  is equal to $d_{\cO} - n$, where $d_{\cO}$ is the physical dimension of $\cO$. }
\beq
\label{f1}
F (1, \ldots , n ) \ = \ \lan 1, \ldots , n | \cO(0) | 0 \ran
 \ .
 \eeq
The external state can in principle contain all different particles in the theory.
At strong coupling, these objects have been considered recently in \cite{Alday:2007he,mz}, where it was found that their calculation is technically equivalent to that of a periodic Wilson loop whose contour is specified by the lightlike momenta of the scattered particles. In distinction to the amplitude calculation at strong coupling \cite{am}, the momenta  do not sum to zero as there is an operator insertion carrying  momentum $-q$; furthermore, 
the sum of the momenta, $\sum_{l=1}^n p_l = q$ is not null, $q^2\neq 0$, and defines the period 
of the Wilson loop contour relevant at  strong coupling. It would be interesting to see whether this Wilson loop/form factor connection
can be extended to weak coupling as was the case for the Wilson loop/amplitude duality \cite{am,dks,bht}. In this paper we will
find some evidence for this at one loop.

At strong coupling, and leading order in $1/\sqrt{\l}$, the form factor calculation is insensitive
to the polarisation of the external particles and, importantly, to the precise choice of the operator, as long as its anomalous dimension is small compared to $\sqrt{\l}$   \cite{mz}. This is similar to a feature of the amplitude calculation of \cite{am}, where the polarisations of the external states are only expected to play a role starting at  one loop in the $1 / \sqrt{\l}$ expansion.
At weak coupling, we have no reason a priori to expect that the form factor will be independent of the choice of the operator.
We will then focus on the simplest class of operators in $\cN=4$ SYM, that of  half-BPS protected operators.

To be specific, we will consider the  operator
$\Tr (\phi_{12}\phi_{12})$ inserted between a state containing two scalars and an arbitrary number of gluons with positive helicity.   The simplest member of this class is constructed with an external state containing  just the two scalars, and is
the Sudakov form factor
$F(q^2) :=  \lan \phi_{12} (p_1) \phi_{12} (p_2) | \Tr \big(\phi_{12}\phi_{12}\big) (0) | 0 \ran$,
where $q:= p_1 + p_2$.
This quantity was  calculated at one and two loops in $\cN=4$ SYM  in a pioneering paper \cite {VN1} by van Neerven,  using an approach based on unitarity applied to Feynman diagrams developed in \cite{VN2}.
Remarkably, the two-loop Sudakov form factor was found to satisfy an iterative relation
very similar to that discovered in \cite{abdk, bds} for the four-point MHV scattering amplitudes.%
\footnote{We review this iterative structure of the form factor in Appendix  \ref{iterform}.}

We begin in Section \ref{suda}  by rederiving the one-loop result for the Sudakov form factor with an elementary application of unitarity.
In Section  \ref{ff1ll} we will move on to consider a more generic class of $n$-point form factors containing $n-2$ positive-helicity gluons in addition to the two scalars. We will first determine their expression at tree level for arbitrary $n$  using recursion relations.
Our result, Equation \eqref{ffmhvtree},  is a holomorphic function of the spinor variables describing the particle momenta, and is a close relative of the Parke-Taylor MHV scattering amplitude. For example, it localises on a complex line in twistor space, as was found in \cite{witten} for the MHV amplitude.

We will then address the one-loop calculation of this class of form factors constructed from a BPS operator and a state containing two scalars and an arbitrary number of gluons.  Our strategy will consist in applying unitarity directly at the level of the form factor, thus bypassing Feynman diagrams, in the same spirit as the unitarity-based approach of \cite{bddk,fusing} for scattering amplitudes. The quantities entering the  form factor cuts  are tree-level form factors and amplitudes. Both are gauge invariant objects,  whose compact expressions we will  recycle inside the loops, thereby obtaining simple one-loop integrands.  The final result for this class of form factors at one loop is remarkably simple --  
see Equation \eqref{final}. It also very reminiscent of the expression for an $n$-point MHV amplitude at one loop, as we will describe.

This close similarity  will serve as an inspiration for  Section  \ref{compp}, where we compare our weak-coupling result \eqref{final} for the form factors to a one-loop calculation of the periodic Wilson loop which, at strong coupling, computes the same quantity. Quite surprisingly, we will find that this periodic Wilson loop calculates the form factor also at weak coupling, or more precisely, the ratio of the one-loop form factor to its tree-level expression. Some issues related to gauge invariance of the prescription will also be discussed.

Finally, we present our conclusions and discuss directions for further research in Section \ref{canc}.
A few appendices complete the paper. In particular, in Appendix \ref{rrr} we write down a BCFW recursion relation
\cite{bcf,bcfw}  for the $n$-point tree-level form factor discussed earlier. In  Appendix \ref{iterform}, we review for the reader's convenience the iterative structure for the Sudakov form factor in $\cN=4$ SYM found  in \cite{VN1}. Appendix \ref{2meapp} contains the expression of the finite two-mass easy box functions, which appear in the form factor and Wilson loop calculations.

\section{The Sudakov  form factor}
\label{suda}
We begin by introducing  the operator
\beq
\label{op}
\cO_{ABCD} \ := \
{\rm Tr} (\phi_{AB} \phi_{CD})  -  {1\over 12}  \eps_{ABCD} {\rm Tr} ( \bar\phi^{LM} \phi_{LM} )
\ ,
\eeq
where $\bar \phi^{AB} := (1/2) \eps^{ABCD} \phi_{CD}$.
The operator in \eqref{op} belongs to the  $\mathbf{20}^\prime$ representation of the  $SU(4)$  $R$-symmetry group  and
is half BPS, i.e.~it has vanishing anomalous dimension. Without loss of generality we will focus in the rest of this paper on its particular component
\beq
\label{op12}
\cO := {\rm Tr} (\phi_{12} \phi_{12} )
\ .
\eeq
In this section we concentrate on
the simplest form factor one can construct using this operator,  namely the two-point or Sudakov form factor,
\beq
\label{ff}
F(q^2)\ := \ \lan \phi_{12} (p_1) \phi_{12} (p_2) | \cO (0) | 0 \ran
\ ,
\eeq
where
\beq
\label{qqq}
q \ := \ p_1 + p_2\ .
\eeq
As mentioned in the Introduction, this form factor in $\cN=4$ SYM was studied at one and two loops in \cite{VN1}, and calculated   in \cite{bds} up to three loops.
Here we will reproduce the one-loop result of \cite{VN1} as a simple application of unitarity. We have also performed a similar calculation at two loops and found  agreement with the result of \cite{VN1}, but we will not discuss it here.

\subsection{The one-loop Sudakov factor from unitarity cuts}

The Sudakov form factor is very simple to compute since, by Lorentz invariance, \eqref{ff} depends only on $q^2$; it can then be determined entirely from its unitarity cut in the $q^2$ channel. In \cite{VN1,VN2} this procedure was  carried out at one and two loops; here we depart from the approach of these two references in that
we apply unitarity directly at the level of the form factor, bypassing Feynman diagrams.

\begin{figure}[h]
\centerline{\includegraphics[height=3.7cm]{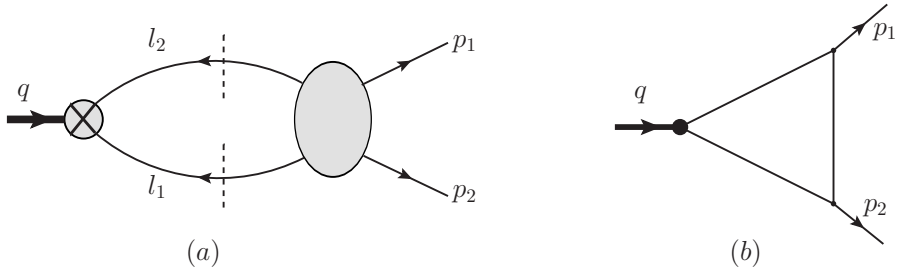} } \caption{\it
In Figure (a)  we show the diagram calculating the cut in the $q^2$-channel of the Sudakov form factor \eqref{ff}. 
The cross denotes a form factor insertion. A second diagram with legs 1 and 2 swapped has to be added and doubles up the result of the first diagram. The result of this cut is given by (twice) a cut one-mass triangle function, depicted in Figure (b).} \label{sudakov}
\end{figure}

The $q^2$-cut of the form factor (i.e.~its discontinuity in the $q^2$-channel) is obtained from the diagram on the left-hand side of Figure \ref{sudakov}, whose expression is%
\footnote{In this and the following formulae we omit a power of the 't Hooft coupling, defined as 
$a:= (g^2 N )  / (16 \pi^2) ( 4 \pi e^{ - \gamma})^\eps $. Note that this is 1/2 the  't Hooft coupling defined  
in \cite{bds}.}
\beq
\label{ffc1}
 \left.
 F^{(1)}(q^2)
\right|_{q^2-\mathrm{cut}}
\,  =  \, 2\int\!\!d{\rm LIPS} (l_1, l_2; q)\  F^{(0)} (l_1, l_2; q) \,
A^{(0)}\big(  \phi_{12} (p_1),  \phi_{12} (p_2),  \phi_{34} (l_1),  \phi_{34} (l_2) \big) \ ,
\eeq
where the Lorentz invariant phase space measure is
\beq
d{\rm LIPS}(l_1, l_2; q) := d^D l_1\, d^D l_2 \, \delta^+ (l_1^2)  \delta^+ (l_2^2) \delta^D (l_1+ l_2  + q)\ ,
\eeq
and $q$ is given in \eqref{qqq}.
The tree-level  component amplitude appearing in \eqref{ffc1},
$ A^{(0)}\big(  \phi_{12} (p_1),  \phi_{12} (p_2),  \phi_{34} (l_1),  \phi_{34} (l_2) \big)$,  
can be extracted from Nair's superamplitude \cite{Nair}
\beq
\label{superMHV}
\cA_{\rm MHV} :=  g^{n-2} \,  (2 \pi)^4 \delta^{(4)}  \big(\sum_{i=1}^n \l_i \lt_i \big) \, \delta^{(8)} \big(\sum_{i=1}^n \l_i \eta_i \big) \, \prod_{i=1}^n {1\over \lan i i+1 \ran}
\ ,
\eeq
where
$\l_{n+1} \equiv \l_1$.
The result is
\beq
\label{ffc2}
A^{(0)}\big(  \phi_{12} (p_1),  \phi_{12} (p_2),  \phi_{34} (l_1),  \phi_{34} (l_2) \big) \ = \ { \lan l_1 l_2 \ran  \lan12\ran\over
\lan l_2 1 \ran  \lan 2 l_1 \ran } \ .
\eeq
The other quantity appearing in \eqref{ffc1}, $F^{(0)}$  is the tree-level expression for the form factor  \eqref{ff}, which is trivially equal to 1. Thus, we get
\beq
\label{zz}
 \left.
 F^{(1)}(q^2)
 \right|_{q^2-\mathrm{cut}}
 =  2\int\!\!d{\rm LIPS} (l_1, l_2; q)\
{\lan12\ran \lan l_1 l_2 \ran \over \lan 2 l_1 \ran \lan l_2 1 \ran }\ = \
-2\, q^2 \int\!\!d{\rm LIPS} (l_1, l_2; q){1\over (l_2 + p_1)^2 }
\ .
\eeq
The expression  in \eqref{zz} represents the cut of a one-mass triangle $T^{\rm 1m} (q^2, \eps)$, depicted in  Figure \ref{sudakov}(b).
It can instantly be lifted to a full loop integral since it depends only on a single kinematic invariant, much in the same spirit as  \cite{Kosower:1999xi}.
Doing so we get
\beq
\label{cc}
F^{(1)} (q^2) \ =  \ 2\,     {\rm Tri}  (q^2, \eps )
\ ,
\eeq
where the function  $ {\rm Tri}  (q^2, \eps ) :=  q^2 T^{\rm 1m} (q^2, \eps)$ is
explicitly evaluated in  \eqref{tri}. This result agrees with that of \cite{VN1}.


\section{Multi-point form factors}
\label{ff1ll}

With the operator $\cO$ introduced in \eqref{op12},    we can now construct an infinite sequence of $n$-point form factors, 
\beq
\label{ff2}
 F(1,\ldots , i_{\phi_{12}} , \ldots, j_{\phi_{12}}, \ldots ,    n; q )\, := \, \lan  g^+(p_1) \cdots  \phi_{12} (p_i) \cdots    
 \phi_{12} (p_j) \cdots g^+ (p_n) | \cO (0) | 0 \ran
\ , 
\eeq
i.e.~we take  matrix elements of $\cO$ between the vacuum and a state containing the same two scalars already appearing in \eqref{ff}, along with $n-2$ positive-helicity gluons. The particular form factor we consider is colour-ordered with respect to the positions of the external particles. However notice that the operator $\cO$  is a colour singlet,  hence the momentum $q$ it carries can be inserted at any position in colour ordering. 
We now present the calculation of \eqref{ff2}, first at tree level and then at one loop.

\subsection{Tree level} 
At tree level, it is easy to calculate the form factor \eqref{ff2}. 
Indeed, we observe that factorisation theorems are  valid also for form factors -- in fact they apply to Green's function in general, see for example \cite{wein} for a discussion. We can then use tree-level factorisation in order to write down a BCFW recursion relation \cite{bcf, bcfw} for form factors. 
 Our result for this quantity is  very simple,%
\footnote{The calculation of \eqref{ffmhv} is presented in Appendix \ref{rrr}. We have also checked our result against  Feynman diagrams in a  few  cases.}  
\beq
\label{ffmhv}
F^{(0)}(1,\ldots , i_{\phi_{12}} , \ldots, j_{\phi_{12}}, \ldots , n;  q )  \ = \   g^{n-2} (2 \pi)^4 \delta^{(4)} ( \sum_{k=1}^n \l_k \lt_k - q) \ F_{\rm MHV}
\ , 
\eeq
where 
\beq 
\label{ffmhvtree}
F_{\rm MHV}\ = \   {\lan ij\ran^2 \over \lan 12\ran \cdots \lan n1\ran }\ . 
\eeq
Here  $p_m := \l_m \lt_m$, and $\sum_{m=1}^n p_m := q$ is the momentum carried by the operator insertion. 
A number of remarks are in order. 

{\bf 1.} The expression in \eqref{ffmhvtree}  is purely holomorphic in the spinor variables of the external particles. In this sense this form factor is the closest off-shell relative of the Parke-Taylor MHV amplitude. We will refer to this form factor as to the ``MHV form factor". Note however that the momenta do not sum to zero. We also note that 
this expression is very reminiscent of the formula for the infinite sequence of Higgs + gluons amplitudes considered in \cite{dgk}. 

{\bf 2.} 
We can easily transform the tree-level form factor   \eqref{ffmhv} to Penrose's twistor space, with the result 
\beq
\label{line} 
F_{\rm MHV} \int\!\!d^4x \ e^{i q x} \, \prod_{m=1}^n \delta^{(2)} ( \mu_m + x \l_m)
\ , 
\eeq
where $(\l_m, \mu_m)$ are the twistor space coordinates of the $m^{\rm th}$ particle. 
As for the case of the MHV amplitude, holomorphicity of the tree-level form factor \eqref{ffmhvtree} ensures that $F_{\rm MHV}$ can be pulled out of the half-Fourier transform.   Equation \eqref{line} shows that 
our MHV form factor is  localised on a line in twistor space, similarly to the MHV amplitude \cite{witten}.

{\bf 3.} We notice that  \eqref{ffmhvtree} satisfies an  auxiliary (helicity) condition
\beq 
{1\over 2} \Big(- \l_m {\partial \over \partial \l_m} +  \lt_m {\partial \over \partial \lt_m}\Big) F(1, \ldots , n)  \ = \
 h_m F(1, \ldots , n) \ , 
\eeq
where $h_m$ is the helicity of the $m^\mathrm{th}$ particle. This relation is just the statement that each external state must transform appropriately under the little group of a massless vector  \cite{witten}.

\subsection{The one-loop MHV form factor from unitarity}

We can now insert  the compact expressions for the tree-level  form factor \eqref{ffmhvtree}  found earlier into 
unitarity cuts, and glue it with  tree-level amplitudes in order  to build cuts of loop form factors. 
Our strategy will consist in calculating the cuts of form factors in all kinematic channels,  and then reconstructing the function which has all the correct cuts, in complete analogy with the approach of \cite{bddk, fusing} for amplitudes. 
The final result will be expressed in terms of the elements of the basis of one-loop scalar integral functions, 
times rational coefficients.%
\footnote{This approach might of course miss purely rational terms, not linked to discontinuities of integral functions. Since we work in $\cN=4$ super Yang-Mills such terms are absent.} 

The operator we have chosen in \eqref{op} is unrenormalised; 
as a consequence, the form factor will contain no ultraviolet-divergent functions. 
At one loop this implies  the absence of bubbles. 
Unlike amplitudes, we will encounter one-mass triangles; however, as we shall see momentarily,  
their purpose will  be that of canceling certain otherwise unwanted infrared divergences in multi-particle channels. 
Indeed, anticipating our story a little, the result of our calculation will consist of a sum of infrared-divergent terms containing only two-particle kinematic invariants, along with a sum over all possible finite parts of two-mass easy  box functions.

We now discuss all possible cuts of the form factor. Here is an outline of the main features of our calculation: 

{\bf 1.} We will perform cuts in  three distinct  channels. The first one is the  $q^2$-channel, where the tree-level form factor entering the cut is a Sudakov form factor. In the second case, which needs to be treated separately, a tree form factor with one additional external particle enters the cut. These two cases give rise to two-mass easy box functions, as well as to triangles. Finally,  in  the most generic channel, the form factor entering the cut expression  has an arbitrary number of legs.
Importantly, in this case we will discover that, after factoring out the tree-level expression of the form factor, 
the integrand will be the same as that of the one-loop MHV amplitude in $\cN=4$ super Yang-Mills, calculated in \cite{bddk}.

{\bf 2.} 
In principle we will have to perform sums over all possible internal states which can run in the loop. We have found that, after performing  the sums over different states when necessary, the cut-integrand becomes the same 
in all channels.  This is in complete analogy  with the calculation of \cite{bddk} of the one-loop 
MHV scattering amplitude in $\cN=4$ SYM. 

{\bf 3.}  The form factor is ordered with respect to the position of the external particles in colour space. 
One may be tempted to think of the form factor as an amplitude with one (or more, if there are more operator insertions)
leg going  off shell, but this picture would not be  true as far as the  colour ordering is concerned since 
the operator insertion is a colour singlet. 
Hence, its position -- the position of the  ``off-shell line" --  does not affect the colour ordering of the external states.  
One must  therefore insert the operator in all possible ways and  sum over the corresponding contributions. 
In practice, this possibility will arise only in the $q^2$-cut diagrams.


\subsubsection{The $q^2$ channel} 

\begin{figure}[h]
\centerline{\includegraphics[height=3cm]{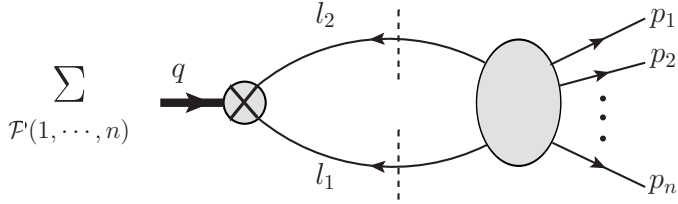} }
\caption{\it The $q^2$-cut of the one-loop form factor. Note that the complete cut is obtained by summing over cyclic permutations of $(1, \ldots , n)$.}
\label{q2-channel}
\end{figure}

We begin by computing the cut  in the $q^2$ channel, where 
\beq
q \, = \, \sum_{k=1}^n p_k \ , 
\eeq 
is the sum  of the momenta of the  external particles.  This cut is represented in Figure \ref{q2-channel},  and is given explicitly by 
\beqa
\label{ffc3} 
&&
\left. 
F^{(1)}(1, \ldots , n; q)  
\right|_{q^2-\mathrm{cut}} 
\ = \ \int\!\!d{\rm LIPS} (l_1, l_2; q)\ \sum_{\cP(1, \ldots , n)} 
 \\ 
&&\Big\{ F^{(0)} (l_1, l_2; q) \ 
A^{(0)} \big( \phi_{34} (l_1),  \phi_{34} (l_2),   g^+(p_1 ),  \ldots ,  \phi_{12} ( p_i),  \ldots  \phi_{12} (p_j),  \ldots,  g^+ (p_n) \big) \Big\}  \ . 
\nonumber
\eeqa
Note that we have included a sum over cyclic permutations of the particles $1, 2, \ldots , n$. This is because the insertion of the operator does not affect the colour ordering, therefore all terms in the sum have the same  ordering in colour space. 

We now move on to describe the ingredients appearing in \eqref{ffc3}.
The tree-level form factor $F^{(0)}$  is equal to 1. The tree amplitude $A^{(0)}$, obtained from \eqref{superMHV},  is 
\beq
A^{(0)} \ = \ {\lan ij \ran^2 \lan l_1 l_2 \ran^2 \over \lan l_2 1 \ran \lan 12 \ran \cdots \lan n l_1 \ran \lan l_1 l_2 \ran } 
\ . 
\eeq
Let us focus on one term in the sum, specifically that with the ordering $l_1, l_2, 1, \ldots, n$ of the legs. The others will be obtained by cyclically permuting $1, \ldots , n$. This term can be written as 
\beqa
\label{ffc4} 
F^{(0)} (1, \ldots , n; q)
\int\!\!d{\rm LIPS} (l_1, l_2; q)\ {\lan l_1 l_2 \ran \lan n1 \ran \over \lan l_2 1 \ran \lan n l_1 \ran} 
   \ , 
\nonumber
\eeqa
where $F^{(0)} (1, \ldots , n; q)$ is  the tree-level MHV form factor \eqref{ffmhvtree}. It is easy to show that 
\beqa
{\lan l_1 l_2 \ran \lan n1 \ran \over \lan l_2 1 \ran \lan n l_1 \ran}  & = &{2 \Big[ (p_1q) (l_1 p_n) - (p_1 l_1) (qp_n) + (p_1p_n) (q l_1) \Big] \over (l_2 + p_1)^2  (l_1 + p_n)^2 } 
\\ \nonumber 
&=&
{ 2 (qp_n) (qp_1) - q^2 (p_1 p_n) \over (l_2 + p_1)^2  (l_1 + p_n)^2} + {(p_1q) \over  (l_2 + p_1)^2} + {(qp_n) \over (l_1 + p_n)^2}
\ , 
\eeqa
where $( p_a p_b) \!:= \!p_a \!\cdot \!p_b$. 
This reduction thus leads to the $q^2$-cut integral of 
a two-mass easy box and two scalar triangles, all of which can be calculated using standard formulae, see \cite{VN1} as well as 
\cite{Brandhuber:2004yw,Bedford:2004py,Quigley:2004pw,Bedford:2004nh,Brandhuber:2005kd} for more recent applications.%
\footnote{The important observation that  dimensionally regularised dispersion integrals are well-defined, and rather simple objects to compute, was made in   \cite{VN2}.} 
The two-mass easy box obtained  here  has massless legs $p_1$ and $p_n$, whereas the momenta of the massive corners are $-q$ and $P_{2 \, n-1} = q-p_n - p_1$,  where 
\beq
\label{pab}
P_{ab} \, := \, p_a + \cdots + p_b\ . 
\eeq
Furthermore, we have two two-mass triangles. The first one has massless leg $p_1$ and massive legs $-q$ and $q-p_1$; for the second we just replace $p_1$ by $p_n$. Notice that by summing over all permutations we will generate twice every possible two-mass triangle, where the massless leg is in turn any one of the momenta of the particles, and one of the massive corners has momentum $-q$. Furthermore, we will produce all possible two-mass easy boxes with massless legs $a-1$ and $a$, and massive legs $-q$ and $P_{a+1 \, a-1}$ for any $a \in (1, \ldots , n)$.

Rather than calculating these dispersion integrals, we will now proceed to inspect other kinematic channels. This will enable us to reconstruct the one-loop form factor from its cuts.

\subsubsection{The cut in a generic   kinematic channel }
\begin{figure}[h]
\centerline{\includegraphics[height=3cm]{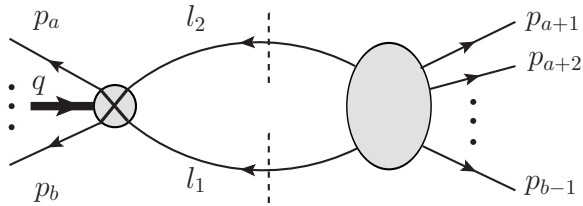} }
\caption{\it The cut of the one-loop form factor  in the $s_{a+1 , b-1}$-channel. }
\label{ab-channel}
\end{figure}
\noindent
Here we discuss the most generic cut, depicted in Figure \ref{ab-channel}.  
There are three possibilities to consider. 
The two scalars can either be emitted both from the form factor; or both from the MHV amplitude; finally, one can be emitted from the form factor, and one from the amplitude. In the latter case, it is necessary to sum over gluons and fermions, which can propagate in the loop. 
Remarkably, the three cases give rise to the same integrand. For this reason, in Figure \ref{ab-channel} we need not distinguish the different types of particles. 

For the sake of definiteness,  we focus on the case where two scalars are emitted from the amplitude. The corresponding expression  is quickly seen to be
\beqa
\label{uno}
\left. {}  \
 F^{(1)} 
\right|_{s_{a+1, b-2 }\!-\!\mathrm{cut}} 
&=& 
\int\!\!d{\rm LIPS} (l_1, l_2; P_{a+1\,   b-1} ) \\
&&
F^{(0)} (-l_2^\phi , - l_1^\phi, b^{g^+},  \ldots, a^{g^+}; q) \, 
\nonumber \\
&&  A^{(0)} \big(l_1^\phi , l_2^\phi, (a+1)^{g^+} \,  \ldots,   i^\phi, \ldots , j^\phi  , \ldots ,  (b-1)^{g^+}\big) \ ,  
\nonumber 
\eeqa
where $P_{ab}$ is defined in \eqref{pab}, and the superscript indicates the particle's species. 

Notice that the particles  emitted from the tree-level form factor fix the colour ordering, hence there is no sum over permutations to be performed in this case. Using the explicit expressions of the tree-level form factor and the amplitude, \eqref{uno} becomes
\beq
\label{gen2pc}
\left. {}  \
 F^{(1)} 
\right|_{s_{a+1, b-2 }\!-\!\mathrm{cut}} 
= 
F^{(0)} 
\int\!\!d{\rm LIPS} (l_1, l_2; s_{a+1,  b-1} ) \ {\lan a \, a+1 \ran \lan l_2\,  l_1 \ran \over \lan a \, l_2 \ran \lan l_2 \, a+1 \ran } \ 
{\lan b-1 \, b \ran \lan l_1\,  l_2 \ran \over \lan b-1 \, l_1 \ran \lan l_1 \, b\ran }
\ , 
\eeq
where $F^{(0)}$ is now the tree-level form factor \eqref{ffmhvtree}. The reader may have recognised that the integrand of \eqref{gen2pc} is the same as that appearing in the two-particle cuts of the one-loop MHV amplitude considered in \cite{bddk}. 
Performing the reduction as in that paper,  one finds that the integrand of  \eqref{gen2pc} can be recast as the sum of four terms,  
\beq
\label{Rsop}
R(b, a+1) + R(b-1, a) - R(b,a) - R(b-1, a+1)\ , 
\eeq
where 
\beqa
\label{R}
R(b,a) & :=  & 
{\lan b \, l_2 \ran  \lan a \, l_1  \ran  \over \lan b \, l_1 \ran \lan a \, l_2\ran } 
\, = \,  {   2 \Big[ (l_1 p_b) (l_2 p_a) + (l_1 p_a) ( l_2 p_b ) - (l_1 l_2 ) ( p_a p_b )    \Big] 
\over 2 (l_1 p_b) \, 2 (l_2 p_a) }
\nonumber \\
&=& 
1+ {(p_bP)\over 2 (p_b\,  l_1) } + { (p_a \, P) \over 2 ( p_a\, l_2)} + { 2 (p_a\, P) (p_b\, P)  - P^2 (p_a\, p_b) \over 2 (p_b \, l_1) 2 (p_a \, l_2) }
\ , 
\eeqa
and we used momentum conservation $l_1 + l_2 + P =0$, where $P$ is the cut momentum ($P^2 = s_{a+1,  b-1}$ in this cut). We also used the cut conditions $l_1^2 = l_2^2 = 0$.

\begin{figure}[ht]
\begin{center}
\scalebox{0.75}{\includegraphics{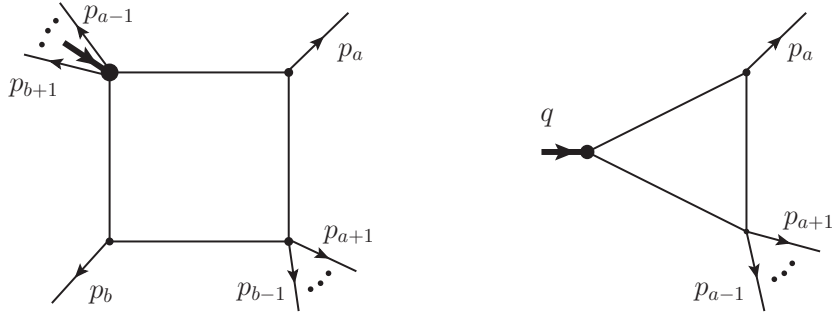}}
\end{center}
\caption{\it
On the left, we represent a two-mass easy box function. The momenta $p_a$ and $p_b$ are null, 
whereas, in general the remaining momenta $P:=  P_{a+1 \, b-1} $ and $Q:=-q + P_{b+1 \, a-1} $ are not null. 
The cases when  either $P^2$ or $Q^2$, or both, are also null, correspond to the 
one-mass and zero-mass boxes, obtained as smooth limits from the expression 
\eqref{2mebst} of the two-mass box function. On the right, we represent a two-mass triangle arising from the cuts considered in Section \ref{2mt}. The thick line represents the momentum carried by the operator. 
}
\label{figure1}
\end{figure}

Equation \eqref{R} shows that  each $R$-function gives rise to a bubble, two triangles and a two-mass easy box function. When all four of the terms in \eqref{Rsop} are present, bubbles and triangles cancel. We will see in the next section that when $b=a$ there are some surviving triangle functions (whereas bubble always cancel among themselves, as anticipated). 

Specifically, the box function arising from $R(b,a)$ has massless legs $a$ and $b$, and massive legs $P_{a+1 \, b-1}$, 
and 
$-q+ P_{b+1 \, a-1}$. Furthermore, it appears here in the corner cut $s_{a+1,  b-1}$, i.e.~the cut momentum legs are those adjacent to the corner momentum $P_{a+1 \, b-1}$. The other $R$-terms  in \eqref{Rsop} will give the same $s_{a+1,  b-1}$-cut of different two-mass easy box functions, where the entries $l$ and $m$ of $R(l,m)$ denote the massless legs of the box. Note that the same box in \eqref{R} will appear in all of its other cuts, with precisely the same coefficient. To see this, notice that the quantity  
$2 (p_a\, P) (p_b\, P)  - P^2 (p_a\, p_b)$ 
is invariant under $P \to P + \alpha \, p_a  + \beta \, p_b$ for any $\alpha$ and $\beta$. 
We remind the reader of the pictorial representations of the  box and triangle functions in 
 Figure \ref{figure1}.

\subsubsection{The cut in a $(q-p_a)^2 $ channel}
\label{2mt}
\begin{figure}[h]
\centerline{\includegraphics[height=3cm]{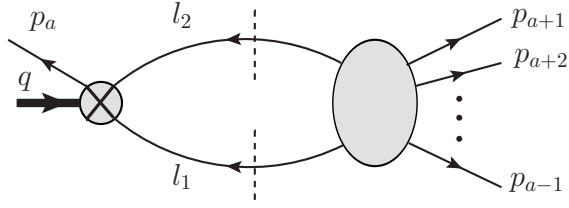} }
\caption{\it The $(q-p_a)^2$-cut of the one-loop form factor. }
\label{qa-channel}
\end{figure}

\noindent
Next we consider a channel where a single particle, of momentum $p_a$,  is emitted from the form factor --
see  Figure \ref{qa-channel}. 
This  $(q-p_a)^2$-channel cut can be regarded as the special $a=b$ case of the  generic $s_{a+1 , b-1}$-channel cut described in the previous section. Note that in the corresponding calculation of the one-loop MHV amplitude, with a single particle of momentum $p_a$ emitted from a vertex,  this cut would be absent since $p_a^2 =0$. However, the momentum flowing from the form factor is here $-q+p_a$, which is not null, and there is a new nontrivial cut to compute. 

The cut integral is given by a formula identical to \eqref{gen2pc} except that now $a=b$. The result for the cut is therefore
\beq
\label{Rsopex}
R(a, a+1) + R(a-1, a) - R(a,a) - R(a-1, a+1)\ , 
\eeq
where, using \eqref{R}, we see that  $R(a,a) = 1$. As a consequence,  bubbles still cancel among the four terms, but there are uncanceled triangles (as well as boxes). Specifically, using the second line of \eqref{R}, it is immediate to see that there is a single surviving triangle, which gives a contribution 
\beq
2\,  { (p_a P) \over 2 ( p_a l_1)} \ , 
\eeq
where in this particular cut, $P = -q+p_a = - P_{a+1  \, a-1}$. This is therefore a two-mass triangle, where the massless leg has momentum $p_a$, and the two massive legs have momentum $P_{a+1 \, a-1}$ and $-q$, respectively. 
We have thus found the same set of two-mass triangles as we found from the $q^2$-cuts discussed earlier.


\subsection{The complete result} 
The next step consists in inspecting the integral functions appearing in all cuts; one then reconstructs the function which 
has the correct cuts in all channels.  An important observation is that the triangle functions (with the precise coefficient they appear with) precisely cancel all infrared-divergent terms in the $q^2$ and $(q-p_a)^2$ multi-particle channels. 
The remaining multi-particle infrared-divergent  terms cancel among all the boxes in the same way as in the calculation of the one-loop MHV amplitude.

After these cancellations are performed, the final result (divided by the tree-level form factor) has a remarkably simple form: it consists of the sum of infrared-divergent terms containing only two-particle invariants made of momenta which are adjacent in colour space, plus a sum of finite parts of two-mass easy box functions, 
where the massless legs are any two of the particle's momenta, all with coefficients equal to one:%
\footnote{A more explicit way to write the sum of finite two-mass easy box functions in \eqref{final} is 
$ \sum_{i=1}^{n} \sum_{r=1}^{[{n\over 2}]-1}
\Bigl(
1 - (1/2) \delta_{n/2  - 1, r}
\Bigr)\,
{\rm Fin}_{n:r;i}^{\rm 2m\,e}$, where the relation to the functions ${\rm  Fin^{2me}}(p_a, p_b, P, Q)$
depicted in Figure \ref{figure1} is obtained by setting 
$p_a=p_{i-1}$, $p_b=p_{i+r}$, and $P=p_i + \cdots +
p_{i+r-1}$.  We give the explicit expression of these finite box functions in Appendix \ref{2meapp}.}
\beq
\label{final}
F^{(1)} (1, \ldots , n; q )   \ = \  F^{(0)} (1, \ldots , n; q ) \, \Big[ - \sum_{l=1}^n  { ( - s_{l l+1})^{- \eps} \over \eps^2}\ + \ 
\sum_{a, b} {\rm  Fin^{2me}}(p_a, p_b, P, Q) \Big] \ . 
\eeq
In the sum on the right-hand side of \eqref{final} the external particles are distributed following cyclic ordering. Note however that the operator insertion, carrying momentum $-q$, does not affect this ordering.

As an example, we list below the finite box functions  appearing in the four-point result: 
$ (1, \{-q\} , 2, \{3, 4\})$ (massless legs 1 and 2);  
$(1, 2, 3,  \{4,  -q\})$, $(1, \{2,  -q\}, 3, 4 )$ (massless legs 1 and 3); 
$(1, \{ 2,3\},  4, \{ -q\})$ (massless legs 1 and 4);  
$(2, \{ -q\}, 3, \{ 4, 1\})$ (massless legs 2 and 3); 
$(2, 3, 4, \{ 1, -q\})$, $( 2, \{ 3, -q\}, 4, 1)$ (massless legs 2 and 4); 
$(3, \{ -q\}, 4, \{ 1,2\})$ (massless legs 3 and 4).  Here we denote by $(a, \{P\}, b, \{Q\})$ a two-mass easy box with massless momenta $p_a$ and $p_b$, and corner momenta $P$ and $Q$.



\section{Comparing form factors to periodic Wilson loops}
\label{compp}

\begin{figure}[t]
\centerline{\includegraphics[height=3cm]{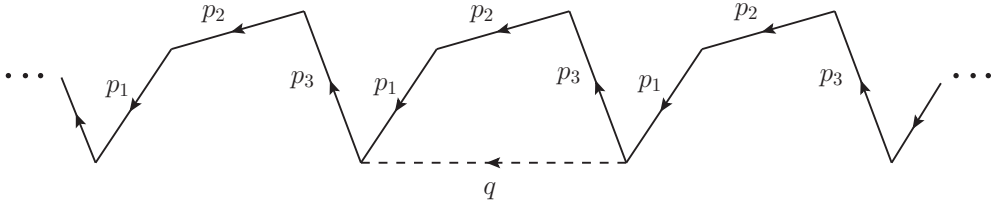} }
\caption{\it Periodic Wilson loop featuring in the duality with form factors. In this case the period  is $q=p_1 + p_2 + p_3$.} \label{periodicWL}
\end{figure}

In  \cite{Alday:2007he,mz}, a prescription to calculate form factors at strong coupling  using  
the AdS/CFT correspondence was proposed. We recall that the form factor calculation,  
as well as the scattering amplitude calculation at strong coupling 
\cite{am, Alday:2007he,  Alday:2009ga,Alday:2009yn,Alday:2009dv,Alday:2010vh},
 are both equivalent to the problem of computing minimal surfaces in AdS space, and hence Wilson loops, but the
boundary conditions in the two cases are quite different. 
For $n$-point amplitudes,  the  boundary (the contour of the Wilson loop)  is the $n$-edged closed polygon 
obtained by joining the lightlike momenta  of the particles following the order induced by the colour structure of the planar 
amplitude.
For $n$-point form factors, the boundary is an infinite periodic sequence of $n$ lightlike segments \cite{Alday:2007he,mz}, see Figure \ref{periodicWL} for an example of  a form factor with three particles, of momenta $p_1$, $p_2$ and $p_3$. The period $q$ is the momentum of the inserted operator, which at strong coupling corresponds to having an additional closed string state inserted on the worldsheet. Therefore, in the form factor case the worldsheet stretches all the way from the boundary at $r=0$ to $r\to \infty$, where $r$ is the radial coordinate of the T-dual AdS space \cite{Alday:2007he,mz}.

It was observed in \cite{dks,bht} that the one-loop calculation of a Wilson loop with the same polygonal contour as in the strong-coupling calculation reproduces the one-loop MHV amplitude, divided by the tree amplitude. 
This was the first manifestation of a new amplitude/Wilson loop duality at weak coupling, which was further checked and studied in \cite{dhks4,dhks5,dhksbum,seven,dhks6,Anastasiou:2009kn, Brandhuber:2009da, DelDuca:2010zg, Brandhuber:2010bj,DelDuca:2010zp, Goncharov:2010jf, Heslop:2010kq}.
An immediate question is therefore whether there is also a 
form factor/periodic Wilson loop duality at weak coupling, i.e.~whether the form factor 
can be equivalently evaluated  at weak coupling from 
a Wilson loop with the same periodic contour found at strong coupling. 
In this section, we calculate the  periodic Wilson loop at one loop, and compare it to  the result \eqref{final} we obtained earlier for the $n$-point form factor using unitarity. 
We will find that the  two are in  agreement. 

The Wilson loop we consider is
\beq
\label{wil}
W[ \cC_n]  \ := \ {\rm Tr} \, \cP \exp \left[ i g\int_{\cC_n} \! d\tau  \ \dot{x}^{\mu} (\tau )A_\mu (x(\tau ))   \right]
\ , 
\eeq
where an example of a periodic contour $\cC_n$ with $n=3$  is depicted in Figure \ref{periodicWL}. 
For the calculation it is convenient to choose the Feynman gauge  and, as discussed in \cite{dks, bht},  this leads at one loop to two classes of diagrams, namely the infrared-divergent cusp diagrams, where a propagator connects two adjacent edges, and those diagrams where the two edges connected by a propagator are not adjacent. Diagrams in the latter class are finite.

The   first evidence of the duality is  straightforward to detect, 
and comes from the infrared-divergent part. The form factor
result \eqref{final} shows that the only  divergent terms contain two-particle kinematic invariants made of momenta that are adjacent in colour ordering, 
\beq
\label{FIR}
F^{(1)} (1, \ldots , n; q ) \left|_{\rm IR}\right.    \  = \ 
F^{(0)} (1, \ldots , n; q)\  \left [ - {1\over \eps^2} \sum_{i=1}^n  \left( {-s_{i i+1}}\right)^{- \eps} \right] 
\ , 
\eeq
with $s_{i i+1} := (p_i + p_{i+1})^2$. 
This contribution  exactly matches the sum of  the one-loop cusp diagrams for the Wilson loop,  
in an identical way as for the matching of the infrared-divergent parts of one-loop MHV amplitudes and Wilson loop \cite{dks,bht}.%
\footnote{We recall that a cusp  contributes a term  $-  (-s_{i i+1})^{-\eps} / \eps^2$ \cite{kk, dks, bht}.  }
As in that case, the dual Wilson loop calculation does not reproduce the tree-level prefactor. 
Note that each period contains $n$ cusps, and we sum once over these $n$ cusps. 
In particular, for the two-point case, the  period contains two cusps which give an identical contribution.  
This is related to the factor of two in the Sudakov form factor \cite{ir1, ir2, ir3, ir4, ir5,  ir6, ir7, ir8}
in \eqref{cc}.

Next, we move to  the finite parts. 
\begin{figure}[t]
\centerline{\includegraphics[height=3cm]{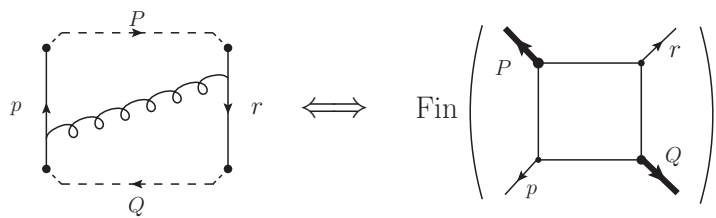} }
\caption{\it We illustrate in this figure the correspondence between Wilson loop diagrams connecting non-adjacent edges and finite parts of two-mass easy box functions. Notice that this correspondence  holds in the Feynman gauge.}
\label{2medual}
\end{figure}
\begin{figure}[t]
\centerline{\includegraphics[height=3cm]{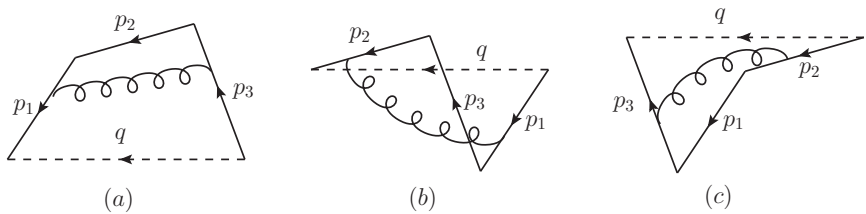} }
\caption{\it One-loop finite Wilson loop diagrams  in the three-point case.  Diagrams (a), (b), (c) are equal to the finite parts of the two-mass easy boxes $(1, 2, 3, \{ -q\})$, $(2, 3, 1, \{ -q\})$, 
and $(2, \{ -q\}, 3, 1)$, respectively.}
\label{WL1loop}
\end{figure}
We found in \eqref{final} that the one-loop form factor contains a sum of finite parts of 
two-mass easy box functions (and one-mass boxes, as special cases of the former). 
It was shown in \cite{bht} that  a Wilson loop diagram  where a propagator connects two non-adjacent edges, 
called $p$ and $r$  on the left-hand side of Figure   \ref{2medual}, is  identical to the 
finite part of a two-mass easy box function with massless legs $p$ and $r$, right-hand side of the same figure.  The massive corners $P$ and $Q$ of the two-mass easy box are  mapped to sums of adjacent momenta  on the Wilson loop side, as shown   
in Figure   \ref{2medual}. This crucial observation puts in one-to-one correspondence the finite boxes appearing in \eqref{final} with all the one-loop Wilson loop diagrams connecting non-adjacent edges in one period. 
For example,  the three-point form factor contains  
the finite parts of  three two-mass-easy boxes 
$(1, 2, 3, \{ -q\})$, $(2, 3, 1, \{ -q\})$, 
and $(2, \{ -q\}, 3, 1)$. 
These functions  are exactly given by the three Wilson loop diagrams depicted in Figure \ref{WL1loop}. 
It is not difficult to see that this mapping is true for general $n$-point form factors.

A few remarks are in order here. 

{\bf 1.} Notice that $q$  is not part of the contour, and is therefore not connected to any other edge by propagators. 

{\bf 2.} 
We draw all finite diagrams where we connect all pairs of non-adjacent legs within a period. 
We can in all cases map these configurations to  finite two-mass easy box functions 
using the correspondence of Figure \ref{2medual}.

\begin{figure}[t]
\centerline{\includegraphics[height=2.8cm]{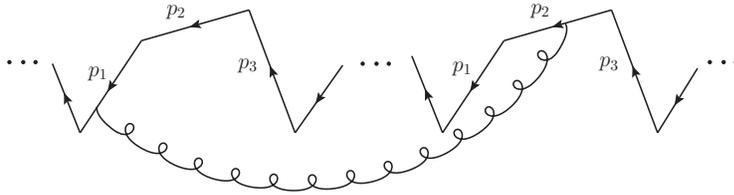} }
\caption{\it In this figure we represent a  one-loop Wilson loop diagram that should not be
included.} 
\label{kk}
\end{figure}

{\bf 3.} In principle one may think that the full  calculation of a  periodic Wilson loop would require a
summation over all translationally  inequivalent diagrams. In particular 
one should also include 
diagrams where a propagator stretches over more than one period, as for example that in Figure \ref{kk}. As we have seen, in order to reproduce the
form factor result these diagrams should not be included. This means
that we need to make a truncation of the periodic Wilson loop and only
consider the diagrams within one period.  Notice that at strong coupling the form factor is calculated 
 by the area of one period \cite{Alday:2007he,mz}.  

{\bf 4.} 
By restricting to one period, we have succeeded in mapping the truncated periodic Wilson loop calculation, which is not obviously  gauge invariant, to the form factor, which is gauge invariant by definition. 
It would be important to explore this issue further, in particular through higher-loop calculations of the periodic Wilson loop. We leave this for future work.

{\bf 5.} An important  difference between the amplitude and the form factor calculations at strong coupling  is the presence in the latter case of additional boundary conditions at $r\to \infty$ due to the operator insertion on the worldsheet. It would be important to understand what modifications of the weak-coupling Wilson loop calculation this entails. This might require the insertion of additional operators compensating the gauge non-invariance of the calculation.%
\footnote{We thank Gregory Korchemsky for discussions on the issue of gauge invariance.}

\section{Conclusions} 
\label{canc}

To summarise, we have calculated form factors  of half-BPS operators in $\cN=4$ super Yang-Mills theory at tree level and one loop. This took advantage of the powerful methods of recursion relations \cite{bcf,bcfw} and unitarity \cite{bddk,fusing}, which may also be applied to form factors.  
The expressions we found for the one-loop form factors with two scalars and an arbitrary number of positive-helicity gluons in particular share salient features with MHV scattering amplitudes. We then found that these
form factors could be straightforwardly derived from periodic Wilson loops, suggesting that there is a new duality between form factors and periodic Wilson loops.

One can also easily generalise the form factor defined in \eqref{ff2} to one containing external states with different particles,
in particular negative-helicity gluons. These should correspond to non-MHV extensions of the form factor we considered.
It would be useful to explore a supersymmetric formulation of such quantities. 
Most importantly, it would be very interesting to consider form factors with insertions of different operators, including non-BPS ones such as the Konishi operator. 

We also found that the tree-level form factors studied here localise in twistor space, and a similar question arises 
as to whether more complicated form factors  (e.g.~with insertions of the Konishi operator)  enjoy similar localisation properties in twistor space.   
We anticipate that form factors similar to \eqref{ff2}, but  where the external states also contain negative-helicity gluons, will be localised on unions of lines as found in \cite{witten} for the tree-level non-MHV scattering amplitudes. 
One would also like to investigate the application of the whole MHV diagram approach to form factors,
much in the same spirit as was done in  \cite{dgk} when considering interactions with massive bosons.

Finally, an obvious important question is to investigate this conjectured form factor/Wilson loop duality at higher loops. 
It seems likely that there will be many further interesting and productive directions to explore as part of applying the techniques and insights obtained from the study of scattering amplitudes to correlation functions more generally.


\vspace{0.4cm}
\section*{Acknowledgements}

It is a pleasure to thank  Lance Dixon, Valeria Gili, Gregory Korchemsky and Sanjaye Ramgoolam  for useful  discussions. 
We would also like to thank Paul Heslop for earlier collaboration on closely related topics, and 
Gregory Korchemsky and Emery Sokatchev for bringing  the paper \cite{VN1} to our attention. 
WJS was supported by a Leverhulme Research Fellowship. 
This work was supported by the STFC under a Rolling Grant  ST/G000565/1.

\appendix

\section{Recursion relations for tree-level form factors}
\label{rrr}

In this appendix, we apply the BCFW recursion relation to calculate
the tree-level form factor defined in \eqref{ff}. The basic inputs 
are three-point amplitudes, plus the three-point MHV and anti-MHV form
factors, 
\bea F^{\rm MHV}_3(i_\phi, j_\phi, k_{g^+}) = {\langle i~j \rangle
\over \langle j~k\rangle \langle k~i \rangle} ~, \quad F^{\overline
{\rm MHV}}_3(i_\phi, j_\phi, k_{g^-}) = {[ i~j ] \over [j~k] [k~i]}
~, \eea
which can be easily derived by using Feynman diagrams.

\begin{figure}[h]
\centerline{\includegraphics[height=2.2cm]{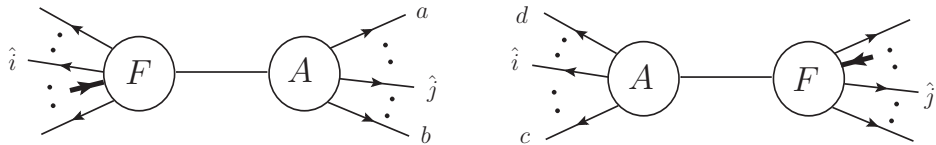} }
\caption{\it BCFW recursion diagrams for tree-level form factors.} \label{BCFW-ff}
\end{figure}

As is standard in the BCFW recursion relation, we use an  $[i,j\rangle$ shift,
$\tilde\lambda_i \rightarrow \tilde\lambda_i - z \tilde\lambda_j$,
$\lambda_j \rightarrow \lambda_j + z \lambda_i$. 
Thus we obtain a one complex parameter family of form factors,  $F(z)$. 
As mentioned earlier, 
factorisation theorems are also valid for form factors; therefore, 
exactly as in the case of scattering amplitudes, by using Cauchy's
theorem  we can calculate the form factor by summing the
residue of the poles from various factorisation channels (we also
require $F(z)\rightarrow0$ as $z\rightarrow\infty$), 
\bea F(0) = \sum_{a,b,h} F_L^h(z\!=\!z_{ab}) {1\over P_{ab}^2}
A_R^{-h}(z\!=\!z_{ab}) + \sum_{c,d,h} A_L^h(z\!=\!z_{cd}) {1\over
P_{cd}^2} F_R^{-h}(z\!=\!z_{cd})~, ~~ \eea
as shown in Figure \ref{BCFW-ff}. Notice that there are two diagrams,
since we can insert the operator either on the left- or on the right-hand
side. 

As an example, we will derive the  expression for the MHV form factor.
Since we require the form factor $F(z)$ to vanish as $z\rightarrow \infty$, we
shift the momenta of two positive-helicity gluons. It is not
difficult to see that the non-vanishing diagrams  are those where the
right-hand side is an anti-MHV three-point amplitude. Without loss of
generality, we calculate the $n$-point form factor $F_n(1_\phi, 2^+,
3_\phi, 4^+, \cdots, n^+)$ by performing  a  $[2,4\rangle$ shift. There are
two non-vanishing channels to consider. The contribution  of the first one is
\bea && F_{n-1}(5^+,\cdots,n^+,1_{\phi_{12}},2^+,\hat P_{\phi_{12}})
{1\over s_{34}} A_3(\hat P_{\phi_{34}}, 3_{\phi_{12}}, 4^+)
\nonumber\\ && = { \langle 1~\hat P\rangle^2 \over \langle 5~6
\rangle \cdots \langle 2~\hat P \rangle \langle \hat P~5 \rangle
}{1\over s_{34}} { [ 3~4 ][ 4~\hat P] \over [\hat P~3 ] } \, , 
\eea
where  $\hat P= P_{34}+z_{34} \lambda_2\tilde\lambda_4, z_{34} =
-\langle 4~3\rangle/\langle2~3\rangle$. The other one is
\bea && F_{n-1}(6^+,\cdots,1_{\phi_{12}},2^+,3_{\phi_{12}},\hat
P'^+)
{1\over s_{45}} A_3(\hat P'^-, 4^+, 5^+) \nonumber\\
&& = { \langle 1~3\rangle^2 \over \langle 6~7 \rangle \cdots \langle
3~\hat P' \rangle \langle \hat P'~6 \rangle }{1\over s_{45}} {
[4~5]^3 \over [\hat P'~4][5~\hat P'] } \, , 
\eea
where $\hat P'= P_{45}+z_{45} \lambda_2\tilde\lambda_4, z_{45} =
-\langle 4~5\rangle/\langle2~5\rangle$. The sum of these two terms
indeed reproduces the MHV form factor result (\ref{ffmhvtree}).
Notice that as an illustration we chose a complicated shift; it would
be much simpler to shift the momenta of two adjacent gluons.

\section{The iterative structure of \cite{VN1} for the two-loop form factor} 
\label{iterform}

In \cite{VN1}, Van Neerven calculated the one- and two-loop contribution to the 
Sudakov form factor $F(q^2, \eps)$, defined in \eqref{ff2}. 
His results are%
\footnote{In this appendix we indicate explicitly the dependence of the form factor on the dimensional regularisation parameter $\eps$.}
\beqa
F^{(1)} (q^2, \eps) &=&2\, {\rm Tri} (q^2, \eps  )\ , 
\\
F^{(2)} (q^2, \eps) & = &4\,  {\rm LT} (q^2, \eps ) + {\rm CT }(q^2, \eps ) 
\ , 
\eeqa
where the one-loop one-mass triangle   ${\rm Tri} (q^2, \eps) $ 
and the two-loop ladder and crossed triangle, $ {\rm LT} (q^2, \eps)$,  ${\rm CT }(q^2, \eps) $ respectively, 
are given by%
\footnote{In the following formulae we actually divide  these functions  by a power of $q^2$ per loop.} 
\beqa
\label{tri}
{\rm Tri } (q^2, \eps )  &=& -  (-q^2)^{-\eps} 
    e^{ \eps \gamma } 
 {\Gamma (1 + \eps ) \Gamma^2 (-\eps) \over  
  \Gamma (1 - 2 \eps) } \\ \nonumber 
  &=& 
  (-q^2)^{-\eps}  \left[- {1\over \eps^2} \, + \,  {\zeta_2\over 2} \,  + \, {7\over 3} \zeta_3 \, \eps \, + \, {47\over 1440} \pi^4 \, \eps^2 
   \, + \, \cO (\eps^3)\right]
     \ , 
 \eeqa
 \beqa
{\rm LT} (q^2, \eps  )  &=&  (-q^2)^{-2\eps}
 e^{ 2  \gamma \eps }     
 \bigg\{  
   {1\over \eps}  \bigg[  {1\over 2 \eps } G (2, 2 ) G_3 (2 + \eps, 1, 1)\\
  &&  - 
      G(2, 1)  \Big[ {1\over \eps}  G_3 ( 2, 1, 1 + \eps)  + G_3 (1, 1, 1) \Big] \bigg]  \bigg\} 
      \nonumber \\ 
      &=& 
     (-q^2)^{-2\eps}  \left[ {1\over 4 \eps^4} + { 5 \pi^2\over 24 \eps^2}  + {29\over 6 \eps}  \zeta_3  + {3\over 32} \pi^4 + \cO (\eps) \right] 
     \ , 
   \nonumber    \\
{\rm CT}(q^2, \eps) &= &  (-q^2)^{-2\eps} \left[{1\over \eps^4} - {\pi^2\over \eps^2}   - {83\over 3 \eps}  \zeta_3  - {59\over 120}  \pi^4 \, + \, \cO (\eps) \right]  
\ , 
\eeqa
where 
\beqa
G(x,y) & =  & 
{ \Gamma ( x + y +  \eps - 2)  \Gamma (
   2 - \eps - x )  \Gamma (  2 - \eps - y )  \over  
     \Gamma (x)  \Gamma (y)  \Gamma ( 4 - x - y - 2 \eps)  }
\ , 
\\ 
G_3 ( x, y, z)  &=& 
{ \Gamma ( 2- x - z - \eps )  \Gamma ( 2- y - z - \eps  ) \Gamma ( 
   -2+x + y + z + \eps  ) 
 \over 
      \Gamma ( x ) \Gamma (y)  \Gamma (4 -x - y - z - 2 \eps  )   }
\ . 
\eeqa
In \cite{VN2}, the functions   ${\rm LT} (q^2, \eps  )$ and ${\rm CT}(q^2, \eps)$ were derived from dimensionally-regulated dispersion integrals; here we present these functions in the form given in \cite{Smirnov:2006ry}. 

In \cite{VN1} it was proved that the Sudakov form factor exponentiates at two loops. 
In formulae, one finds 
\beq
\label{sopraa}
 F^{(2)} (q^2, \eps) - {1\over 2} \Big( F^{(1)} (q^2, \eps)\Big)^2 \ = \ (-q^2)^{- 2 \eps} \left[ 
 {\zeta_2 \over \eps^2} + {\zeta_3 \over \eps} + \cO (\eps)\right] \ . 
 \eeq
 We can recast \eqref{sopraa} in an ABDK/BDS form, namely 
 \beq
 F^{(2)} (q^2, \eps) - {1\over 2} \Big( F^{(1)} (q^2, \eps)\Big)^2 \ = \  f^{(2)}_{\rm FF} (\eps) F^{(1)} (q^2, 2 \eps) +  C^{(2)}_{\rm FF} +  \cO (\eps)\ , 
 \eeq
where $f_{\rm FF}^{(2)}  (\eps) = \tilde{f}_0 + \tilde{f}_1 \eps + \tilde{f}_2 \eps^2$. We then find
\beq
\tilde{f}_0 = - 2 \zeta_2\ , \qquad \tilde{f}_1 = - 2 \zeta_3 \ . 
\eeq
We also find a condition relating $f_2$ and $C^{(2)}_{\rm FF} $, namely 
\beq
C^{(2)}_{\rm FF}   = { \tilde{f}_2\over 2} +  {\pi^4\over 18}  
\ . 
\eeq
On the other hand, the four-point MHV amplitude (divided by the tree-level amplitude) satisfies \cite{abdk, bds}
 \beq
 \label{su}
M^{(2)}  ( \eps) - {1\over 2} \Big( M^{(1)} ( \eps)\Big)^2 \ = \  f^{(2)}  (\eps) M^{(1)} (2 \eps) +  C^{(2)} +  \cO (\eps)\ , 
 \eeq
with 
$f^{(2)}  (\eps)= f_0 + \eps f_1 +\eps^2 f_2$ and  
\beq
\label{dd2}
f_0 = -\zeta_2 =  {\tilde{f}_0 \over 2} \ , \qquad f_1 = - \zeta_{3} = {\tilde{f}_1 \over 2} \ . 
\eeq
The factor of 1/2 in the  result \eqref{dd2} is a matter of convention -- it can be understood once one recalls that $f_0$ and $f_1$ are written in a convention where the 't Hooft coupling $a_{\rm BDS}$ is twice as that used in the present paper (as well as in \cite{VN1}). Indeed, inspecting \eqref{su} one quickly realises that 
the combination  $a f^{(2)}$ must be independent of any conventions used to define the coupling, since the left-hand side $a^2 M^{(2)}$ is clearly convention independent.

\section{The two-mass easy box function}
\label{2meapp}
A compact   form of the finite part of a two-mass easy box function containing only four dilogarithms was first 
derived in \cite{Duplancic:2000sk}, and then  found independently in \cite{Brandhuber:2004yw}  in the context of 
MHV diagrams, where an analytic proof of its equivalence  with the conventional expression of  
e.g.~\cite{Bern:1993kr}  was given. 
Expressing the two-mass easy box  as a function of the kinematic invariants 
$s := (P+p)^2$,  $t := (P+q)^2 $ and $P^2$,  $Q^2$, with $p+q+P+Q=0$, 
its finite part is
\beq
\label{2mebst}
  {\rm  Fin^{2me}} (s,t,P^2, Q^2) \, = \, 
 \Li(1-aP^2)\, + \, \Li(1-aQ^2)  \, -\,  \Li(1-as)
\,  -\,  \Li(1-at)
\ , 
\eeq
where 
\beq
\label{adef} 
a \ = \
\frac{P^2+Q^2-s-t}{P^2Q^2-st} \ . 
\eeq


\end{document}